\documentclass[%
 reprint,
 amsmath,amssymb,
 aps,
]{revtex4-1}

\usepackage{gensymb}
\usepackage{graphicx}
\usepackage{dcolumn}
\usepackage{bm}
\usepackage{hyperref}

\usepackage{verbatim}

\usepackage[dvipsnames,svgnames,x11names]{xcolor}

\usepackage{dsfont}
\usepackage[Symbol]{upgreek}
\usepackage{physics}

\usepackage{subfigure}



\begin{document}

\preprint{APS/123-QED}

\title{Unconventional superfluidity of superconductivity on Penrose lattice}

\author{Yu-Bo Liu}
\thanks{These two authors contributed equally to this work.}
\author{Zi-Yan Shao}
\thanks{These two authors contributed equally to this work.}
\author{Ye Cao}
\email{ycao@bit.edu.cn}

\author{Fan Yang}
\email{yangfanblg@bit.edu.cn}

\affiliation{School of Physics, Beijing Institute of Technology, Beijing 100081, China}

\date{\today}

\begin{abstract}
We theoretically investigate the gap function, superfluid density and the transition temperature of the superconductivity (SC) on semi-periodic Penrose lattice, where an attractive Hubbard model is adopted as an example. Firstly, we clarify that the gap function, density of states and superfluid density are all positively correlate to the extended degree of single particle states around the Fermi energy. Secondly, we identify that the paramagnetic component of the superfluid density does not decay to zero in the thermodynamic limit, which is completely different from the periodic system. The difference between the diamagnetic and paramagnetic currents keeps stable with whatever scaling, which is consistent with recent experimental results that although the superfluid density is lower than that of the periodic system, the system has bulk SC.  Thirdly, we find that both the superfluid density and SC transition temperature can be boosted with the increase of disorder strength, which should be general to quasicrystal but unusual to periodic systems, reflecting the interplay between the underlying geometry and disorder.
\end{abstract}

\maketitle

\section{Introduction}

The study of superconductivity (SC) in a system without translational symmetry
has aroused widespread interest since the 1960s \citep{Hasse1968,Comberg1974,Bergmann1976}.
As a special kind of such systems, quasicrystal (QC) has particularly
attracted much attention \citep{Goldman1993} since it was synthesized
because of its long-range order and special symmeties \citep{Shechtman1984},
as a result of which, the electronic structure on a QC is essentially
different from that on either a crystal or a disordered lattice, making
it possess various exotic quantum states \citep{Andrade2015,Autti2018,Giergiel2019,Huang2018,Huang2019,Kogo2017,Kraus2012,Lang2012,Longhi2019,Otsuki2016,Shaginyan2013,Takemori2015,Takemura2015,Thiem2015,Tsunetsugu1991-1,Tsunetsugu1991-2,Watanabe2016,Wessel2003,Yamamoto1995,Guo2020,Xu2022}.
However, it was not until 2018 that the convincing evidence for the
emergence of bulk SC in a synthesized Al-Zn-Mg
QC was discovered \citep{Kamiya2018}. This groundbreaking work together
with explorations in previous ternary QCs \citep{Wong1987,Wagner1988}
and crystalline approximants \citep{Deguchi2015}, have set off an
upsurge in the study of QC SC.

In one dimension, the interplay between an external potential driven
quasidisorder and pairing mechanism in a periodic lattice has been
deeply investigated in both condensed matter and ultracold atomic
systems \citep{Tezuka2010,Tezuka2013,DeGottardi2013,Cai2013,J.Wang2016,Y.Wang2016,Cao2016,Ghadimi2017,Roati2008}.
In a higher dimension, acompanying in the growth of experimental technology,
scholars have paid attention to SC in mesoscopic quasiperiodic systems,
such as quasiperiodic networks of ordinary superconducting wires \citep{Roati2008,Behrooz1986,Gordon1986,Springer1987,Nori1987,Nori1988,Niu1989}
and quasiperiodic pinning arrays in ordinary superconductors \citep{Misko2005,Misko2006,Kemmler2006,Silhanek2006,Misko2010}.
In recent years, more concerns has arisen in the study of SC in
microscopic quasiperiodic systems, setting up the research of QC
superconductors in many aspects, the definition of topological
invariance \citep{Fulga2016}, real-space distribution of Cooper pairs
and its crossover from extended to localized \citep{Sakai2017,Ara=0000FAjo2019},
Fulde-Ferrell-Larkin-Ovchinnikov mechanism \citep{Sakai2019}, Cooper
instability \citep{Liu2022-1}, fast numerical calculation method
\citep{Nagai2020}, the classification of the pairing symmetries
\citep{Cao2020}, and interaction driven Mott transition \citep{Sakai2022}, to name a few. Some physical properties of quasiperiodic
superconductors have also been calculated numerically to compare with
experimental results and further understand QC SC \citep{Takemori2020}.
In addition, the study of SC in extrinsic QC constructed
from twisted bilayer materials have made important breakthroughs, high-temperature
and high-angular-momentum topological superconductors, which cannot
emerge in periodic systems, are proposed in twisted bilayer cuprates
\citep{Can2021} and graphene \citep{Liu2022-2} respectively.

In our previous work \citep{Liu2022-1}, we analytically proved that the Cooper instability for infinitesimal attractive interaction, which is well known for periodic lattices, also holds for the QC once the density of state (DOS) on the Fermi level is finite and nonzero, and that our mean-field (MF) study revealed a BCS-like pairing phase on the QC either at zero or finite temperature, carrying finite superfluid density. However, there are still general problems to be discussed about QC SC. Firstly, what is the main factor affecting the strength of pairing at zero temperature? Secondly, in order to understand the different behaviors of superfluid density in QCs and periodic systems, we are naturally curious about the scaling behavior of the internal structure, i.e. the diamagnetic and paramagnetic components, of superfluid density. Thirdly, the geometry of QC can be regarded as a kind of disorder to a certain extent. What are the consequences of the interplay between the QC geometry and disorder in SC?

To address these problems,  we calculate the gap function and superfluid density based on BdG theory and obtain the following results: (i) There are positive correlations between the extended degree of the single particle states around Fermi energy and density of states (DOS), pairing strength as well as the superfluid density. (ii) We analyze the scaling behaviors of paramagnetic and diamagnetic superfluid densities by calculating them on system up to $\sim8\times 10^4$ lattice sites. Unlike the periodic system, the paramagnetic current does not decay to zero in the thermodynamic limit, which leads to a relatively small total superfluid density compared with the periodic system. Considering that the superfluid density is usually positively correlated with the $T_c$, this result gives the reason why the $T_c$ decreases when the system transform to QC from the periodic structure in the experiment~\cite{Kamiya2018}. Nevertheless, the superfluid density, i.e. the difference between the diamagnetic and paramagnetic components, remains basically steady when the system size increases, confirming that the system has SC in the thermodynamic limit.  (iii) We also reveal the effect of nontrivial interplay between the geometry of QCs and the external disorder. Our research has yielded an exotic discovery: when states around the Fermi level are more heavily influenced by the geometric modulation of QCs and consequently exhibit greater localization, the SC can be amplified by increasing the strength of external disorder.

The rest of the paper is organized as follows. In Sec.~\ref{Sec:model}, we describe the details of the model we adopted. In Sec.~\ref{Sec:gap_function}, we calculate the gap function and extended degree of the states and present the positive correlation between them. In Sec.~\ref{Sec:superfluid_density}, we reveal the internal structure of superfluid density with two components, and for the first time clarify that there will be residual paramagnetic superfluid density in the thermodynamic limit. Positive correlation is once again determined between superfluid density and the extended degree. In addition, we find an exotic character that both the superfluid density and the transition temperature could be boosted with the increase of disorder when the states around the Fermi level is relatively localized. These properties should be unique to QCs. Finally, Sec.~\ref{Sec:conclusion} is devoted to the conclusions.

\section{Model} \label{Sec:model}
\begin{figure}[htbp]
\centering
\includegraphics[width=0.8\linewidth]{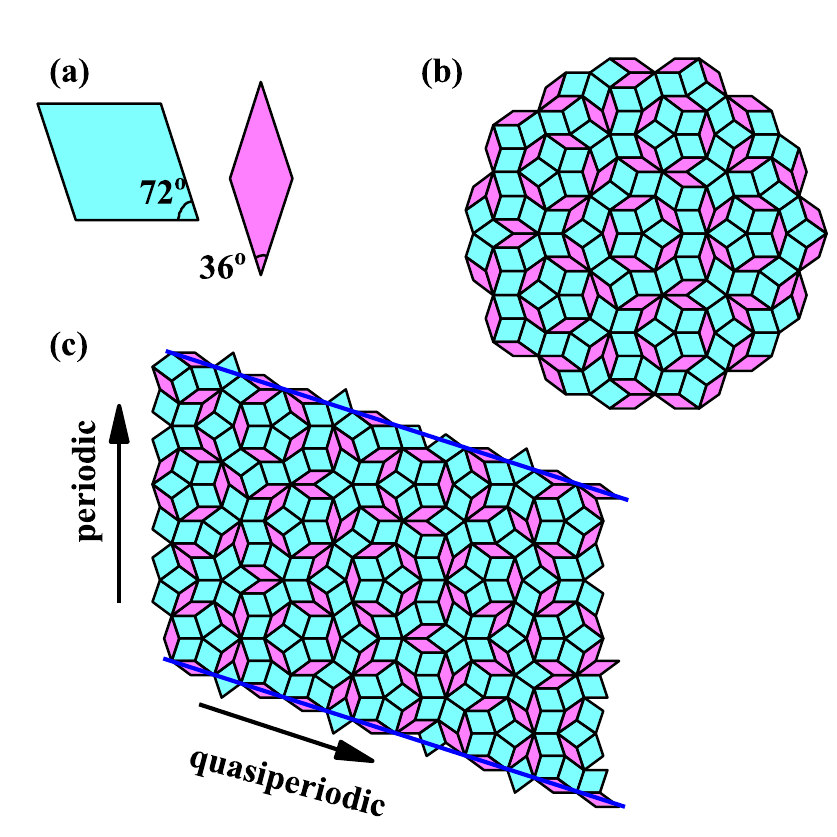}
\caption{\label{Fig:sppl}(Color online) Two kinds of rhombuses (a) are used to generate Penrose tiling (b) and SPPL (c) which is periodic in $y$ direction but remains its original quasiperiodic structure in $-36\degree$ direction. The size of periodic direction of SPPL, i.e. the number of rhombuses (sharing edges perpendicular to the periodic direction with their neighbors) along the periodic direction, can only be some specific value, e.g. 16 in (c). Whilst the size of quasiperiodic direction of SPPL can be arbitrary.}
\end{figure}

The system studied in this work is an attractive Hubbard model on a semi-periodic Penrose lattice~\cite{Tsunetsugu1991-2} (SPPL, as shown in fig.~\ref{Fig:sppl}).
This tiling is a kind of modified Penrose lattice which is periodic in $y$ direction but remains its original feature in $-36\degree$ direction with arbitrary length.
The real-space Hamiltonian is written as
\begin{equation}
\label{Eq:tb-real-space}
\mathcal{H} = -\sum_{ ij\sigma} t_{ij}c^{\dag}_{i\sigma}c_{j\sigma}  -\mu\sum_{i\sigma}n_{i\sigma}-U\sum_{i}n_{i\uparrow}n_{i\downarrow},
\end{equation}
where $c_{i\sigma}$ annihilates an electron at site $i$ with spin $\sigma$, $n_{i\sigma}$ is the electron-number operator, and $\mu$ denotes the chemical potential. The atoms are placed at the center of each rhombus of the Penrose tiling and electrons can only tunnel between the rhombuses sharing one edge with each other. The indices $i$ and $j$ in Eq.~\eqref{Eq:tb-real-space} indicate the centers of the rhombus. In the calculation, we set $t_{\langle ij \rangle}=1$ as the energy unit.
Along the periodic direction, we adopt periodic boundary condition, while along the quasiperiodic direction, an open boundary condition is put to use.

The tight-binding Hamiltonian in Eq.~\eqref{Eq:tb-real-space} can be diagonalized as
\begin{equation}
\label{Eq:tb-state-sapce}
\mathcal{H}_{\mathrm{TB}} = \sum_{m\sigma}\tilde{\epsilon}_m c_{m\sigma}^{\dagger}c_{m\sigma},
\end{equation}
where $m$ denotes a single particle energy level, $\tilde{\epsilon}_m=\epsilon_m-\mu$ is the energy shift relative to Fermi energy and $c_{m\sigma}=\sum_{i}\xi_{im}c_{i\sigma}$.

\section{Gap Function} \label{Sec:gap_function}
In our previous work~\cite{Liu2022-1}, we have clarified that the Cooper instability still holds in QC. Unlike the plane wave in the periodic system, the single particle states in the QC are not completely extended and the extended degree is closely related to doping. We naturally want to ask the relationship between the gap function and extended degree.

Through the same treatment in Ref.~\cite{Liu2022-1} and taking into account the constraint of Anderson's theorem, we arrive at the gap function
\begin{equation} \label{Eq:gap_equation}
\Delta_{m} = \frac{U}{2N} \sum_{n} \frac{ f_{mn}\Delta_{n}}{\sqrt{\tilde{\varepsilon}_{n}^2+\abs{\Delta_{n}}^2}} \tanh\left(\frac{\sqrt{\tilde{\varepsilon}_{n}^2+\abs{\Delta_{n}}^2}}{2k_B T}\right),
\end{equation}
where
\begin{equation}
\begin{aligned}
f_{mn}&=N\sum_{i}\xi^{2}_{im}\xi^2_{in},\\
\Delta_m&=-\frac{U}{N}\sum_{n}f_{mn}\expval{c_{n\downarrow}c_{n\uparrow}}.
\end{aligned}
\end{equation}
Previous study~\cite{Liu2022-1} has shown that the superconductivity in this model can be well described by the BCS theory, and we have found that the overall pairings come out almost unanimously in amplitude. Therefore, ignoring the differences in amplitude, the gap function~\eqref{Eq:gap_equation} evolves into
\begin{equation} \label{eq_gap_equation_cosntant_Delta}
\Delta = \frac{U}{2N} \sum_{n}\frac{\Delta}{\sqrt{\tilde{\varepsilon}_{n}^2+\Delta^2}}\tanh\left(\frac{\sqrt{\tilde{\varepsilon}_{n}^2+\Delta^2}}{2k_B T}\right).
\end{equation}

\begin{figure}[htbp]
\centering
\includegraphics[width=0.98\linewidth]{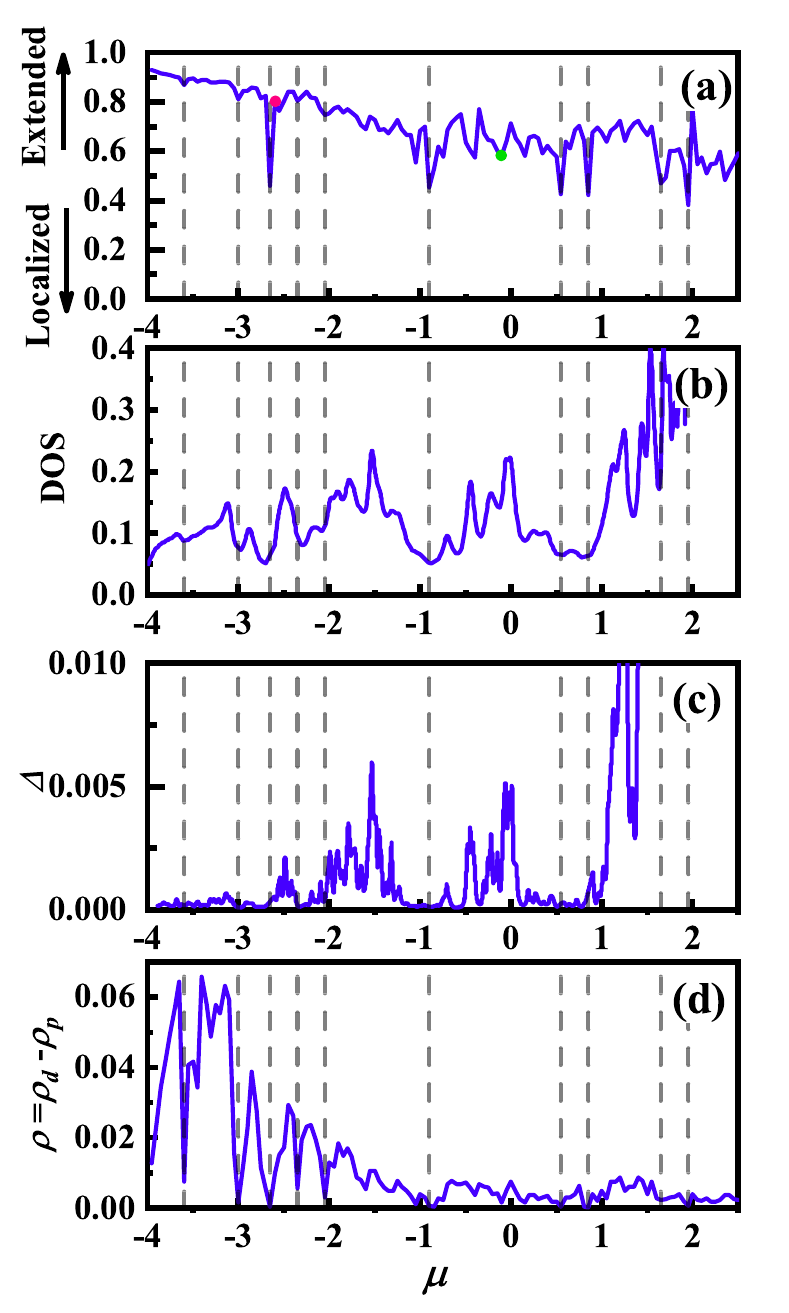}
\caption{\label{Fig:extend_vs_others}(Color online) Extended degree (a) $y=-\log_{N}(x)$ defined in Eq.~\eqref{Eq:extended_degree}, DOS (b), pairing amplitude $\Delta$ (c), and superfluid density (d) as functions of chemical potential $\mu$. We acquire the result from a SPPL with 81509 sites to better reflect the properties in the thermodynamic limit. The Hubbard interaction strength applied in (c) and (d) is $U=0.2$.}
\end{figure}

In order to indicate to what extent the single particle state is extended, we introduce $y=-\log_{N}x$ to mark the extended degree of a state, where $N$ is site number of the system and
\begin{equation}
\label{Eq:extended_degree}
x_{m} = \frac{\sum_{i}\abs{\xi_{im}}^{2p}}{\left(\sum_{i}\abs{\xi_{im}}^2\right)^p}
\end{equation}
is the $2p$-norm of a state defined by H. Tsunetsugu et. al in Ref.~\cite{Tsunetsugu1986}. According to definitions mentioned above, the greater the extended degree of the single particle state, the smaller the value in the Eq.~\eqref{Eq:extended_degree}. Here we adopt the case of $p=2$. In extreme cases, $y=0$ and $y=1$ correspond to the fully localized and extended states respectively. The doping- (i.e. the chemical potential $\mu$-) dependence of the extended degree is shown in Fig.~\ref{Fig:extend_vs_others}(a).

 The $\mu$- dependences of the DOS at Fermi level and the pairing gap amplitude $\Delta$ at zero temperature are shown in Fig.~\ref{Fig:extend_vs_others}(b) and (c), in comparison with the extended degree shown in Fig.~\ref{Fig:extend_vs_others}(a). As shown in Fig.~\ref{Fig:extend_vs_others}(a)-(c), there are many extrema for these quantities. The typical local minima marked with the gray dashed lines in Fig.~\ref{Fig:extend_vs_others}(a) manifest that the extended degree suddenly drops at specifical chemical potentials. Interestingly, the local extrema of the DOS and $\Delta$ shown in Fig.~\ref{Fig:extend_vs_others}(b) and (c) almost coincide with those of the extended degree shown in Fig.~\ref{Fig:extend_vs_others}(a). These results show that the more localized the single particle state around the Fermi energy is, the smaller the DOS is and the weaker the pairing gap amplitude is.

\section{Superfluid Density} \label{Sec:superfluid_density}
An important quantity to depict the superfluid properties of a 2D Hubbard model is the superfluid density tensor, which characterizes system dependent Meissner effect and relates to KT-Nelson criterion of Berezinskii-Kosterlitz-Thouless transition.

The superfluid density can be derived from the linear response to the external magnetic field.
Using the Peierls substitution, the presence of a uniform vector potential corresponds to a change just in the hopping term. In our pervious works~\cite{Cao2020, Liu2022-1}, we have provided the derivation of the real-space current operator in detail.
Here, in order to better describe the behavior in thermodynamic limit, only the current along the quasiperiodic direction (see Fig.~\ref{Fig:sppl}(c)) is considered. Therefore, we remove the direction label in the following. For a small $\mathbf{A}$ in the quasiperiodic direction, the gauge independent current operator in the same direction is derived as
\begin{equation}
\begin{aligned}
\hat{j} & = \frac{\mathrm{i}}{N} \sum_{ij\sigma} t_{ij}r_{ij} c^{\dag}_{i\sigma}c_{j\sigma} - \frac{1}{N} \sum_{ij\sigma} t_{ij}r_{ij}^2 A c^{\dag}_{i\sigma}c_{j\sigma} \\
& = \hat{j}_{p} - \hat{j}_{d},
\end{aligned}
\end{equation}
where $r_{ij}$ is the projection of the vector from site $i$ to site $j$ in the direction of $\mathbf{A}$. 

The current can be divided into two parts, a paramagnetic current
\begin{equation}
\hat{j}_{p}  = \frac{\mathrm{i}}{N} \sum_{mn\sigma} J_{mn} c^{\dag}_{m\sigma}c_{n\sigma},
\end{equation}
where $J_{mn}=\sum_{ij}t_{ij}r_{ij}\xi_{im}\xi_{jn}$, whose direction is the same to $\mathbf{A}$,
and a diamagnetic current
\begin{equation}
\hat{j}_{d} = - \frac{1}{N} \sum_{mn\sigma} \left(\sum_{ij}t_{ij}r_{ij}^2A\xi_{im}\xi_{jn}\right) c^{\dag}_{m\sigma}c_{n\sigma},
\end{equation}
whose direction is antiparallel to $\mathbf{A}$.
The superfluid density $\rho_s\equiv\lim_{\mathbf{A}\to 0}\frac{-\langle\mathbf{A}|\hat{J}[\mathbf{A}]|\mathbf{A}\rangle}{A}$ is also separated into two parts
\begin{equation}
j_{p} = \expval{\hat{j}_{p}} = \rho_{p}A,
\end{equation}
and
\begin{equation}
j_{d} = \expval{\hat{j}_{d}} = \rho_{d}A,
\end{equation}
where $\rho_{p}$ and $\rho_{d}$ are the paramagnetic and diamagnetic superfluid densities respectively.
The $\hat{j}_{d}$ is linearly dependent on $A$, therefore, $\rho_{d}$, can be obtained as the mean value of $\hat{j}_d/A$ in the state without vector potential,
\begin{equation}
\begin{aligned}
\rho_{d} = & -\frac{1}{N} \sum_{m} \left(\sum_{ij}t_{ij}r_{ij}^2\xi^{\ast}_{im}\xi_{jm}\right) \\ & \times \left[1-\frac{\tilde{\varepsilon}_{m}}{\sqrt{\tilde{\varepsilon}_{m}^2+\Delta^2}}\tanh\left(\frac{\beta\sqrt{\tilde{\varepsilon}_{m}^2+\Delta^2}}{2}\right)\right].
\end{aligned}
\end{equation}

The $\hat{j}_{p}$ does not depend on vector potential explicitly, it should disappear with vanishing $A$. According to the linear response theory
\begin{equation}
\rho_{p} = \int_{-\infty}^{\infty} -\mathrm{i} \theta(t) \expval{\comm{A(t)}{B(0)}} \mathrm{e}^{\mathrm{i}\left(\omega+\mathrm{i}0^{+}\right)t} \dd{t}, \omega=0,
\end{equation}
where $A=\hat{j}_{p}$ and $B=H_{\mathrm{TB}}(\mathbf{A})-H_{\mathrm{TB}}(\mathbf{A}=0)=-N\hat{j}_{p}$.
There followed the Lehmann's representation of $\rho_{p}$ at zero temperature
\begin{equation} \label{Eq:rho_p}
\begin{aligned}
\rho_{p} & = 2N \sum_{\mathrm{Ex}} \frac{\abs{\matrixel**{\mathrm{Ex}}{\hat{j}_{p}}{\mathrm{G}}}^2}{E_{\mathrm{Ex}}-E_{\mathrm{G}}} \\
& = \frac{2}{N} \sum_{mn} \frac{\abs{J_{mn}}^2\left(u_{m}v_{n}-v_{m}u_{n}\right)^2}{\sqrt{\tilde{\varepsilon}_{m}^2+\Delta^2}+\sqrt{\tilde{\varepsilon}_{n}^2+\Delta^2}},
\end{aligned}
\end{equation}
where $|\text{G}\rangle$ denotes the BCS ground state and $|\text{Ex}\rangle$
indexes an eigenstate of the BdG Hamiltonian. To deduce the Eq.~\eqref{Eq:rho_p},
we have used the Bogoliubov transformations $ \gamma^{\dag}_{m\uparrow} = u_m c^{\dag}_{m\uparrow} - v^{\ast}_m c_{m\downarrow}$ and $\gamma_{m\downarrow} = v_m c^{\dag}_{m\uparrow} + u^{\ast}_m c_{m\downarrow}$, employing which the paramagnetic current operator is written as
\begin{equation}
\label{Eq:j_p_Bogoliubov}
j_p = \mathrm{i}\sum_{mn} J_{mn} (u_m v_n -u_n v_m) (\gamma_{m\uparrow}^{\dagger} \gamma_{n\downarrow}^{\dagger} - \gamma_{m\downarrow}^{\dagger} \gamma_{n\uparrow}^{\dagger})+\cdots,
\end{equation}
where $\cdots$ represents all the terms that vanish in the off-diagonal average, i.e. $\langle \mathrm{Ex}| \cdots |\mathrm{G}\rangle$ = 0.
For the periodic system, the two energy indexes $m,n$ in the above formula are replaced by two momentum indexes $\mathbf{k},\mathbf{k}'$ respectively and $J_{\mathbf{k}\mathbf{k}^{\prime}}=\sum_{ij}t_{ij}r_{ij,x}\exp(-\mathrm{i}\mathbf{k}\cdot\mathbf{r}_i)\exp(\mathrm{i}\mathbf{k}^{\prime}\cdot\mathbf{r}_j)$.
\begin{figure}[htbp]
\centering
\includegraphics[width=0.98\linewidth]{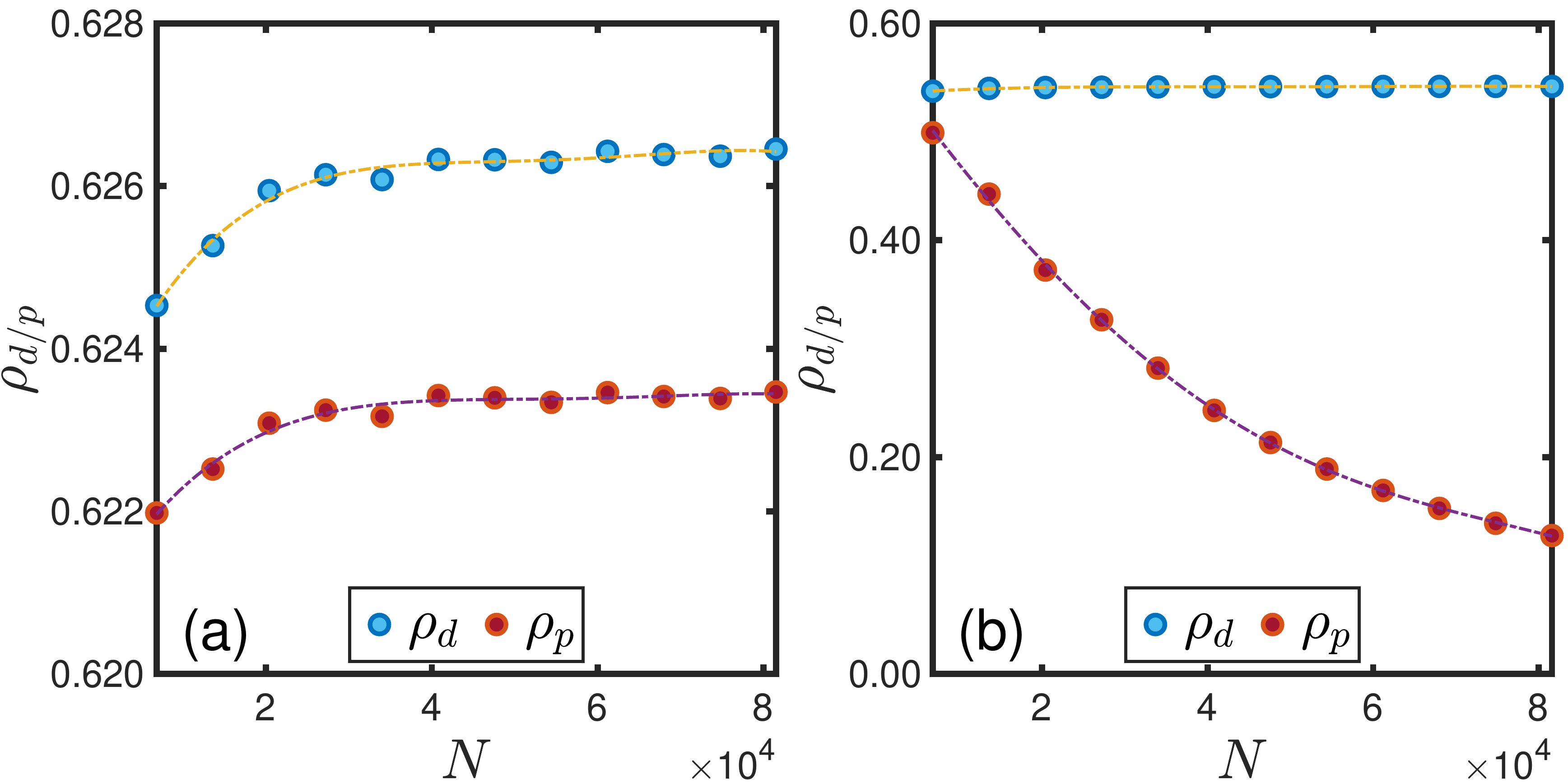}
\caption{\label{Fig:scaling}(Color online) Scaling behaviors of $\rho_{d}$ and $\rho_{p}$ on the SPPL (a) and on the square lattice with open boundary condition (b). The same set of parameters are input on the two lattices: the chemical potential $\mu=1.1$ and the order parameter $\Delta=0.005$. In both systems, the $\rho_{d}$ does not change obviously with the system size scale.  However, when the system size scale increases, the $\rho_{p}$ in the square lattice gradually vanishes, but that in the SPPL remains finite.}
\end{figure}

In order to address the exotic behavior of superfluid in QC, we compare the scaling behavior of superfluid densities in SPPL and square lattice in Fig.~\ref{Fig:scaling}.
For the sake of avoiding the finite size effect as much as possible, we calculate the system from $\sim1\times 10^4$ to $\sim8\times 10^4$ lattice sites, which has almost reached the limit of exact diagonalization.
In both of the systems, the diamagnetic component of superfluid density does not change significantly when the lattice size increases. Meanwhile, when the system size increases, the paramagnetic component of superfluid density gradually disappears as expected in the periodic system, but it remains finite in the SPPL due to the absence of translational symmetry in QC.

The aforementioned phenomenon can be understood as follows. In periodic systems, the $\rho_{p}$ vanishes as the coherence factor $(u_{\mathbf{k}}v_{\mathbf{k}}-v_{\mathbf{k}}u_{\mathbf{k}})=0$ due to the translational symmetry which ensures that $J_{\mathbf{k} \mathbf{k}'}$ is diagonal. However, in QC, the off-diagonal excitations in Eq.~\eqref{Eq:j_p_Bogoliubov} do not disappear, so the summation $\sum_{mn}$ is two-folded here. As a result, $\rho_{p}$ is composed of many non-negative terms, so it is positive in itself.

We also reveal the relationship between the value of zero-temperature superfluid density and the extended degree of the single particle eigenstates around Fermi energy, as shown in Fig.~\ref{Fig:extend_vs_others}(a) and (d).  In local perspective, it can be seen that the degree of localization of the single particle states is inversely related to the superfluid density, that is, the difference between the diamagnetic  and the paramagnetic part. This is reflected in the consistent local extrema of extended degree and superfluid density, we witness that Fig.~\ref{Fig:extend_vs_others}(a) and (d) share the same dips and peaks, especially for the typical dips indicated by dashed lines. It is worth noting that this relationship is in keeping with the aforementioned tie-up between the extended degree and order parameter $\Delta$, indicating that the extended degree, as a unique property of QCs, plays a fundamental role in the formation of SC and other long-range orders. The positive correlation between extended degree and superfluid density is consistent with our intuitive understanding that the more localized the states forming SC are, the more unfavorable it is for the supercurrent to flow somewhere far away.
According to Fig.~\ref{Fig:extend_vs_others}, we speculate scrupulously that the superfluid density would turn to zero in thermodynamic limit at specific chemical potential where the calculated superfluid density is relatively small.  Then, we can put our conjecture:  as clarified in Ref.~[8], many states on the Penrose lattice are actually critical, showing power-law decaying wave function and hence power-law decaying conductance. For the corresponding doping levels of these states, our conjecture is that the superfluid density would also decay in power law with the size of the system. This conjecture is to be verified by large scale numerical calculations on much larger lattices.
\begin{figure}[htbp]
	\centering
	\includegraphics[width=0.98\linewidth]{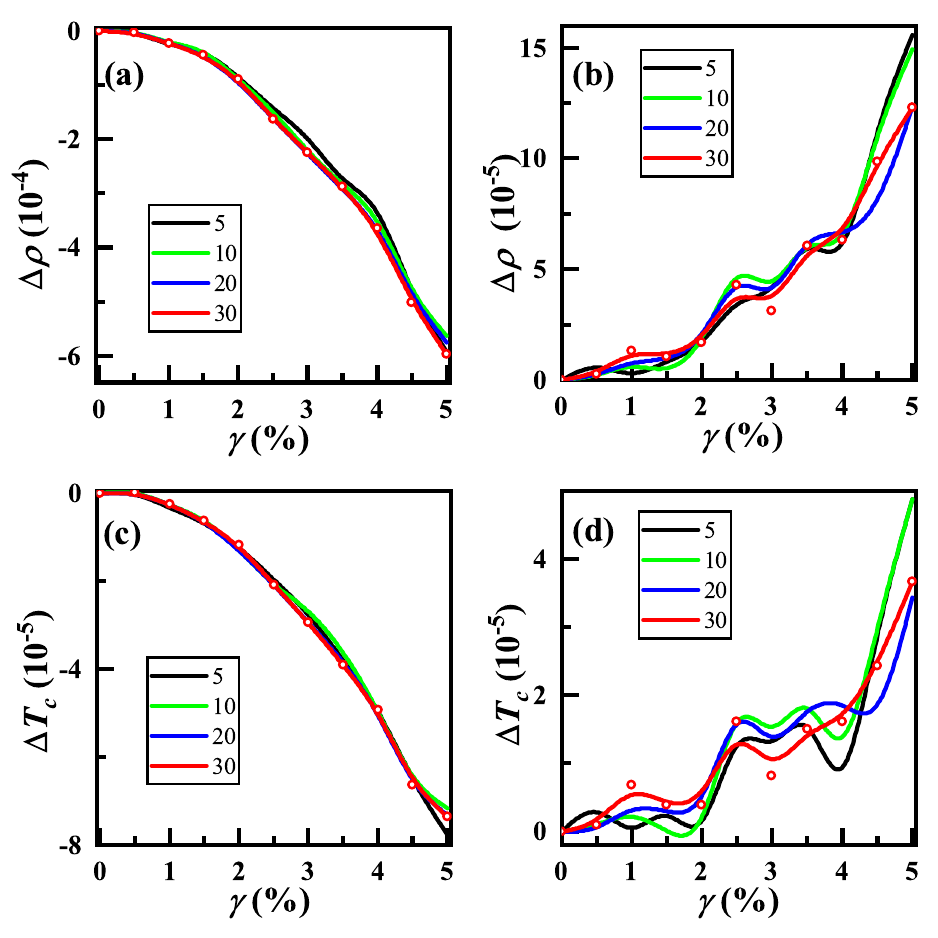}
	\caption{\label{Fig:disorder}(Color online) The deviation of superfluid density (a, b) and $T_c$ (c, d) from their respective clean values as functions of disorder strength. The results in left (a, c) and right (b, d) panel describe the SC formed by extended ($\mu = -2.55$, as marked with red circle in Fig.~\ref{Fig:extend_vs_others}(a)) and localized ($\mu = -0.11$, as addressed with green circle in Fig.~\ref{Fig:extend_vs_others}(a)]) electron states respectively.  
	We adopt the commonly used uniformly distributed disorder, i.e. $H_{W} = \sum_{i\sigma}h_i n_{i\sigma}$, where the local magnetic field $h_i\in[-\gamma, \gamma]$ and $\gamma$ is presented as a percentage of $t_{\langle i j\rangle}$. The curves in different colors correspond to different number of disorder realizations as marked in corresponding legends. On the left, the disorder is detrimental to SC in terms of both superfluid density and $T_c$, however, the disorder is advantageous to both of them on the right.}
\end{figure}

Another interesting question is how the quasi-periodic property of the QC interplays with disorder in the aspect of SC. It's already known that in the aspect of normal-state conductance, a slight disorder imposed on the QC system will not suppress the conductance but enhances it instead. Here in the aspect of SC, we have known that any disorder in the periodic system will essentially harm the formation of SC order. However, how the disorder interplays with the QC structure in this aspect is a question worth studying. In our work, we adopt the on-site disordered potential, the Hamiltonian of which is presented in the caption of Fig.~\ref{Fig:disorder}. According to the numerical simulation of different doping conditions, we find that for the SC formed by the relatively extended states around the Fermi energy, for instance the states around $\mu = -2.55$ marked with red circle in Fig.~\ref{Fig:extend_vs_others}(a), that is, the states closer to those in the periodic system, disorder will tend to damage the SC, as shown in Fig.~\ref{Fig:disorder}(a) and (c). Nevertheless, when the single particle states around Fermi energy are more localized, e.g. the states around $\mu = -0.11$ labeled with green circle in Fig.~\ref{Fig:extend_vs_others}(a), both the superfluid density and the superconducting transition temperature increase with the strengthening of the emerging disorder, as can be clearly seen in Fig.~\ref{Fig:disorder}(b) and (d), which indicates that the SC is fond of disorder at this time.

 This exotic character can be understood as follows. In the perfect periodic case, the superfluidity of the system gives expression to the rigidity of the phase against an applied transverse vector potential. As a result, there
 is only a diamagnetic response of the current, which is the paradigm of Meissner effect in clean superconductors. In QCs, when SC is formed by relatively extended states, the situation should be similar to that of clean superconductors.  In weak disorder, the response of BdG quasiparticles to the applied magnetic field makes an appearance, while the phase of the order parameter remains unchanged. This response leads to a budding paramagnetic current, thereby suppressing the total suppercurrent. However, when the geometric modulation of the QCs is more obvious for the electron states forming SC, the paramagnetic current originated from quasiparticle excitations is significant in itself and thus cannot be ignored (see Eq.~\eqref{Eq:j_p_Bogoliubov} and Fig.~\ref{Fig:scaling}(a)). In this scenario, the emerging disorder is incompatible with the eigenmodes of BdG quasiparticles. The disorder would scatter the zero temperature excitations caused by the geometry of QCs, thus inhibiting the paramagnetic current in the embryonic stage of disorder, therefore boosting the overall suppercurrent.

\section{Conclusion} \label{Sec:conclusion}
In summary, we have studied the superfluidity of superconductivity on a semi-periodic Penrose lattice driven by an attractive Hubbard model. We conclude that (i) the gap function and superfluid density are positively correlated with the extended degree of the single particle states around the Fermi energy. (ii) at zero temperature, the scaling behavior of the paramagnetic part of the superfluid density approaches a finite value rather than zero, which is completely different from the situation in the periodic system. This behavior causes the superfluid density, i.e. the difference between diamagnetic and paramagnetic component, to be smaller than that in the periodic system in the thermodynamic limit, thus indirectly reducing the superconducting critical temperature, which is consistent with the results observed in the experiment~\cite{Kamiya2018}. (iii) superconductivity formed by more localized states near the Fermi energy shows a preference for disorder, which is reflected in the fact that when weak disorder is applied, both the superfluid density and the transition temperature are boosted with the increase of disorder.

\begin{acknowledgments}
This research was supported by the NSFC under the Grant Nos. 12074031, 12234016, 12174024, 11704029, and 11674025.
\end{acknowledgments}


\end{document}